\documentclass[journal,10pt,draftclsnofoot,onecolumn]{IEEEtran}
\ifCLASSINFOpdf
\else
\fi
\usepackage{graphicx}
\usepackage{amsmath}
\usepackage{amsfonts}
\usepackage{amssymb}
\usepackage{psfrag}
\usepackage{algorithm}
\usepackage{algorithmic}
\usepackage[tight]{subfigure}
\usepackage{multirow}
\usepackage{epsfig}
\usepackage{epstopdf}
\usepackage{citesort}

\usepackage{wasysym}
\usepackage{array}

\begin{document}
%
\title{Implementation of Privacy-preserving SimRank over Distributed Information Network}


\author{\IEEEauthorblockN{Yu-Wei Chu$^1$, Chih-Hua Tai$^2$, Ming-Syan Chen$^{1,3}$ and Philip S.
Yu$^4$}\\
\IEEEauthorblockA{$^1$Dept. of EE, National Taiwan University, Taipei, Taiwan\\
$^2$Dept. of CSIE, National Taipei University, New Taipei, Taiwan\\
$^3$Research Center of Info. Tech. Innovation, Academia Sinica, Taipei, Taiwan\\
$^4$Dept. of CS, University of Illinois at Chicago, IL, USA\\
hanatai@mail.ntpu.edu.tw; mschen@citi.sinica.edu.tw;
psyu@cs.uic.edu}}


%



\maketitle

\begin{abstract}
Information network analysis has drawn a lot attention in recent
years. Among all the aspects of network analysis, similarity measure
of nodes has been shown useful in many applications, such as
clustering, link prediction and community identification, to name a
few. As linkage data in a large network is inherently sparse, it is
noted that collecting more data can improve the quality of
similarity measure. This gives different parties a motivation to
cooperate. In this paper, we address the problem of link-based
similarity measure of nodes in an information network distributed
over different parties. Concerning the data privacy, we propose a
privacy-preserving SimRank protocol based on fully-homomorphic
encryption to provide cryptographic protection for the links.

\end{abstract}


%
\IEEEpeerreviewmaketitle

\section{Introduction}

Real network data usually consist of objects of different types,
forming a so called heterogeneous network. A heterogeneous network
contains abundance of information, which gives rise to a great
interest in its analysis such as clustering, classification,
centrality measure, object ranking and pattern mining
\cite{getoor2005link,sun2009ranking,ji2011ranking}, to name a few.
Among all the aspects of heterogeneous network analysis, measuring
the similarity between nodes is one of the most important
problems{}. Answering how similar nodes are is essential in many
applications, such as clustering, link prediction and community
identification
\cite{cai2005mining,liben2007link,jarvis1973clustering}.

According to the evaluation of \cite{maguitman2006algorithmic},
link-based similarity measures
\cite{jeh2002simrank,koren2007measuring,leicht2006vertex}{} well
conform with human judgement. In the work of
\cite{leicht2006vertex}, the similarity of two nodes can be
understood as a weighted count of the number of all-length paths
between the nodes. In \cite{koren2007measuring}, the similarity can
be regarded as the probability that a node A reaches a node B in a
restricted random walk manner. Both the measures proposed in
\cite{leicht2006vertex} and \cite{koren2007measuring} are applied on
undirected networks. SimRank \cite{jeh2002simrank} defines the
similarity score of two nodes A and B as the expected meeting
distance, i.e., the expected distance that two random surfers,
starting at node A and node B respectively, travel before they meet
at the same node. Among all the link-based similarity measures,
SimRank\cite{jeh2002simrank} is one of the most influential
approaches \cite{li2010fast}{} for following reasons. First, SimRank
similarity score is independent to the clustering approaches, and
thus can be applied to various kinds of clustering methods. Second,
SimRank can be applied on both directed and undirected graphs, which
give a great flexibility in analyzing heterogeneous networks.
Furthermore, SimRank can work on any graphs that have
object-to-object relations, without the need of domain knowledge.

\begin{figure}
\centering $\begin{array}{cc}
    \includegraphics[width=1.5in]{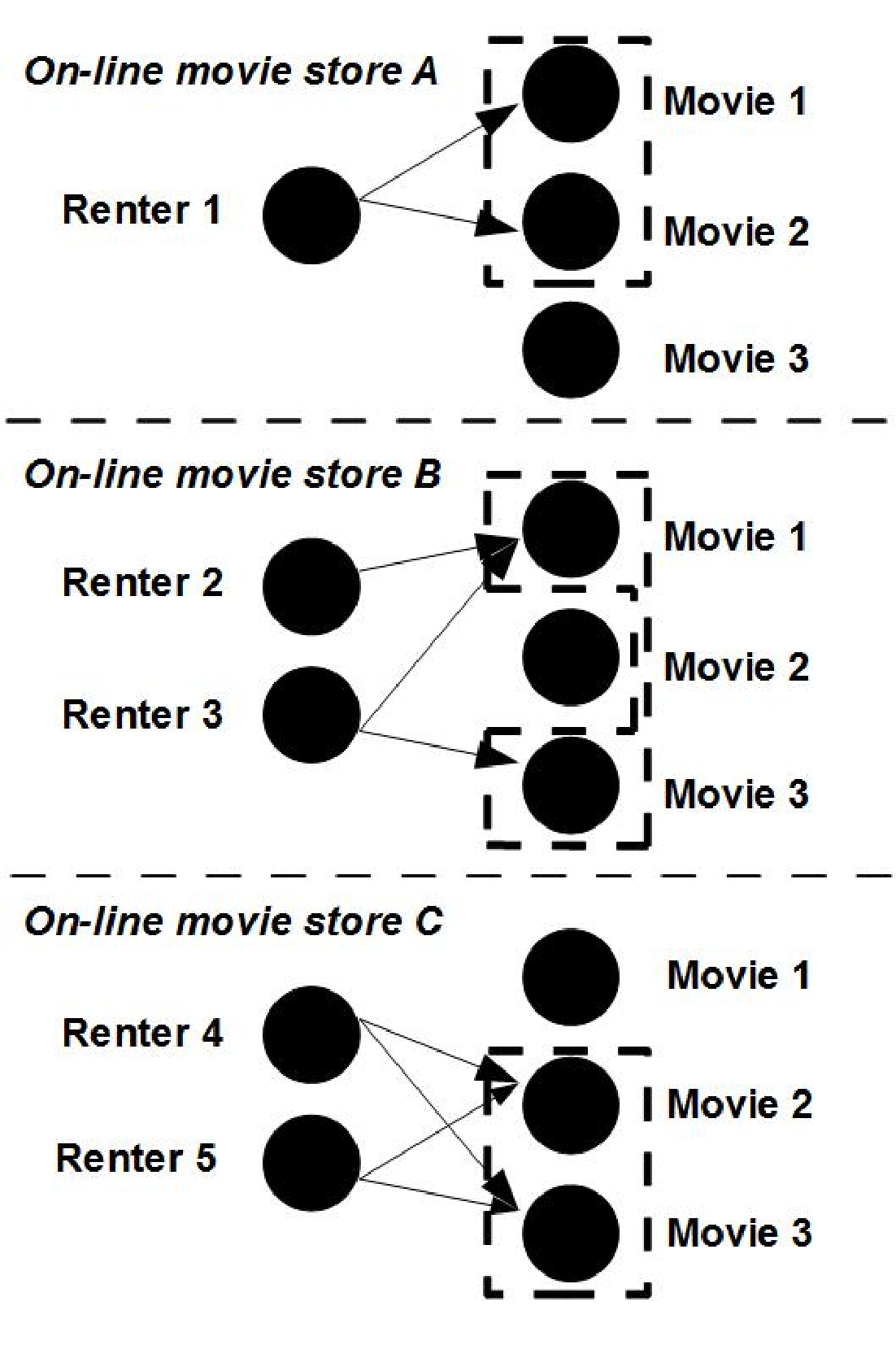} &
    \includegraphics[width=1.5in]{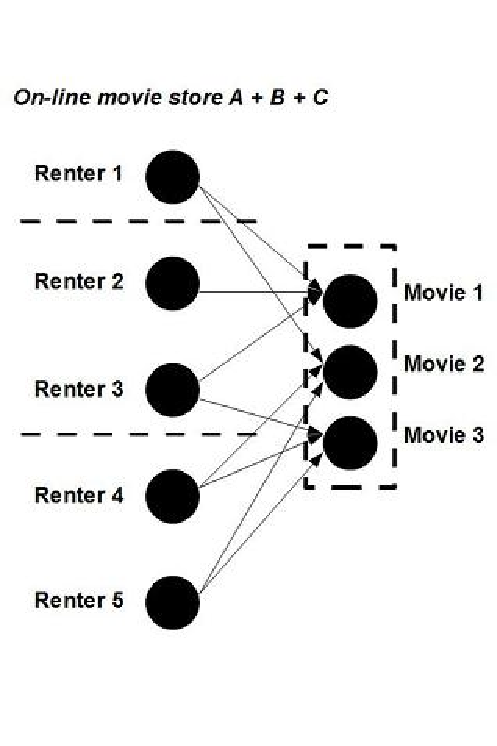} \\
(a) & (b) %
\end{array}$
\caption{The rental data of on-line movie stores A, B and C.}
\label{coComputationBenifit} \vspace{-3mm}
\end{figure}

In reality, however, data are often distributed among different
parties. For example, Figure~\ref{coComputationBenifit}(a) shows the
rental data in three on-line movie stores. Consider that an on-line
movie store wants to build a recommendation system for its
customers. Based on the rental records, SimRank can provide a good
similarity measure for clustering the movies of similar types.
However, the clustering result may still not be good enough since
the amount of rental records is also a dominant factor. Combining
the rental data from different stores for clustering, as shown in
Figure~\ref{coComputationBenifit}(b), surely can improve the
performance of movie recommendation systems. This gives the on-line
stores a motivation to cooperate.

Whenever different parties cooperate, data privacy is an important
concern. In the above example, the on-line movie stores may not be
willing to reveal their network data to each other since the links
indicating the rental information, i.e., customer-rent-movie, are
usually one of the most valuable assets for a company. Revealing the
rental information of a company may allow its competitor to be able
to analyze its customer behavior and use the information against the
company. Therefore, how to cooperatively compute SimRank similarity
score without exposing any sensitive information, i.e., links
between objects, becomes a crucial problem. To the best of our
knowledge, there is not yet any studies addressing this problem.

It is challenging, under the constraint of link protection, to
calculate the link-based similarity of nodes in a network
distributed over different parties. As the similarity calculation is
based on the link information of the entire network, it is essential
to securely integrate the distributed data into a synthesized
graph.{} For example, according to the similarity definition of
SimRank, two nodes are seen as similar if the neighbor nodes they
connect to are themselves similar to each other. As the result, the
entire graph structure needs to be considered when computing
similarity scores between nodes. However, synthesizing graphs
without exchanging any link information seems daunting. Without the
knowledge of the entire graph structure, we can't even correctly
identify the neighbors of a certain node. How to build a combined
graph having access to the information we need and avoid revealing
each party's link information in the same time is an issue needed to
be solved.

Specifically, in this paper we take the link-based similarity
measure defined by SimRank \cite{jeh2002simrank} and address the
problem of privacy-preserving similarity measure in a joint
bipartite network consisting of data from two or more parties. For
such a problem, it has the challenges on (1) how each party can
guarantee its link protection in a way that allows the construction
of a structure containing the information involved for similarity
measure, and (2) how the similarity is calculated on such a
structure. To tackle these challenges, we propose a new protocol
called privacy-preserving SimRank (PP-SimRank) based on
fully-homomorphic encryption FHE
\cite{gentry2009fully,gentry2011implementing}. FHE takes only 1 or 0
as inputs, outputs different ciphertexts with random noise
introduced, and has both the addition and multiplication homomorphic
properties. Therefore, the XOR- and AND- operations on a plaintext
are equivalent to the addition and multiplication on the
corresponding ciphertext. PP-SimRank then utilizes these properties
to construct \textit{virtual} joint networks in ciphertext field. By
extending the similarity definition to the virtual networks,
PP-SimRank is able to calculate SimRank scores without knowing the
link information in the bipartite network of each party. We show
that PP-SimRank can protect the link information in a semi-honest
model, where a semi-honest security model restricts all parties to
faithfully follow their specified protocol, but allows the parties
to record all the intermediate messages for analysis. In addition,
to overcome the constraint of FHE taking only 1 or 0 as inputs,
PP-SimRank converts the SimRank scores into binary representations
and carries out the score computation in ciphertext field with
binary arithmetic operations. That is, we demonstrate an application
of FHE in practice.

Generally, the contributions of this paper include:

\begin{enumerate}
\item {}We are the first to address the problem of link-based similarity measure co-computation over distributed information network.
\item We propose a privacy-preserving protocol, PP-SimRank, to securely compute SimRank similarity score over distributed data. PP-SimRank is the first privacy-preserving protocol that focuses on link-based similarity measure co-computation.
\item We carry out the implementation of basic arithmetic operations, i.e., addition, subtraction, multiplication and division, in the ciphertext field under fully-homomorphic encryption, demonstrate an application of fully-homomorphic encryption scheme and show its potential in privacy-preserving data mining. \newline
\end{enumerate}

\vspace{-3mm} The rest of this paper is organized as follows.
Section~\ref{Sec:2} formally defines our problem and provides the
background knowledge to our work. Section~\ref{Sec:3} describes our
Privacy-Preserving SimRank protocol in details, and
Section~\ref{Sec:4} discusses the implementation issues. We then
analyze the complexity and security of the proposed PP-SimRank in
Section~\ref{Sec:5}, extend PP-SimRank to multi-party scenario in
Section~\ref{Sec:6}, and conclude this paper in Section~\ref{Sec:7}.

\section{Preliminary}
\label{Sec:2} In this section, we formally define the problem and
provide the necessary background. We first formulate the problem in
Section \ref{Sec:ProblemFormulation}, and give a brief review of
SimRank in Section \ref{Sec:SimRank overview}.

\subsection{Problem Formulation}
\label{Sec:ProblemFormulation}

In this paper, we address the problem of privacy-preserving
computation of link-based similarity measure over horizontally
partitioned information network in different parties. Our target is
to protect the link information and achieve the computation of
SimRank score simultaneously. For simplicity, we first focus on
two-party scenario, and then show the extension to general cases.

In a two-party scenario, two parties called Alice and Bob hold
information networks $G^A =(V^A, U, E^A)$ and $G^B=(V^B, U, E^B)$,
respectively. Without loss of generality, we assume that $V^A$ and
$V^B$ are disjoint vertex sets, and $U$ is a vertex set known to
both Alice and Bob. The parties wish to cooperatively learn the
SimRank similarity scores $S(u_i, u_j)$ of all  vertex pairs $\{u_i,
u_j\} \in U$ using their joint data, i.e., $G = (V^A \cup V^B, U,
E^A \cup E^B)$. For privacy concern, the problem is how to construct
the virtual information network $G$ without revealing $E^A$ and
$E^B$ and how to perform the SimRank computation on the virtual
information network $G$.

For example, consider that Alice and Bob are two on-line movie
stores. Alice and Bob have different customer groups, $V^A$ and
$V^B$, respectively, and the same movie set $U$. A rental record is
then a link connecting $v_i$ with $u_j$. When both Alice and Bob
regard their own rental records as sensitive data, the problem is
that, how Alice and Bob can construct a virtual graph $G=(V^A \cup
V^B, U, E^A \cup E^B)$ without revealing {}$E^A=\{(v^A_i, u_j)\}$
and {}$E^B=\{(v^B_i, u_j)\}$ to each other, and learn the SimRank
similarity score $S(u_i, u_j)$ of all the movie pairs to improve
their movie recommendation system.

For this problem, ideally, Alice and Bob would find a trusted third
party to perform all the needed calculations for them. That is,
Alice and Bob would securely transmit their data to this trusted
third party. Let the third party compute the movies' pairwise
similarity scores using SimRank and send the result information back
to both Alice and Bob. However, such independent third parties are
very hard to find or do not exist in the real world.

Instead of counting on a third party, we proposed a
privacy-preserving protocol, PP-SimRank, based on fully-homomorphic
encryption (FHE) \cite{gentry2009fully,gentry2011implementing} to
protect the link information. The fully-homomorphic encryption
method takes the plaintexts of binary form, i.e., $\{0, 1\}$, as
inputs and outputs the ciphertexts in the form of big integers.
Given the same plaintext and the same public key for encryption, FHE
will output different ciphertext (big integer) with random noise,
i.e., there is a negligibly small probability
\cite{gentry2011implementing,kemeny1966introduction} that Enc($1$)
$=$ Enc($1$). Therefore, FHE is semantic secure against chosen
plaintext attacks and is suitable for link encryption. In addition,
FHE has the following two properties.
\\

\noindent \textbf{PROPERTY 1.} Addition-homomorphic
{\arraycolsep=2pt
\begin{equation}
\begin{array}{rcl}
    c_{1} + c_{2} = Enc(m_{1}) + Enc(m_{2}) = Enc(m_{1} {}\oplus m_{2}),\\
\end{array}
\label{Eq:AddHomomorphic}
\end{equation}

\noindent \textbf{PROPERTY 2.} Multiplication-homorphic:
\begin{equation}
\begin{array}{rcl}
    c_{1} \times c_{2} = Enc(m_{1}) \times Enc(m_{2}) = Enc(m_{1} \wedge m_{2}),\\
\end{array}
\label{Eq:MulHomomorphic}
\end{equation}
} \vspace{-3mm}

\noindent where the two ciphertexts $c_1 = Enc(m_1$) and $c_2 =
Enc(m_2$) represent the encryption outcomes of the plaintexts $m_1$
and $m_2$ respectively. Accordingly, the {}addition and
multiplication operations on the ciphertexts are equivalent to the
XOR- and AND-operations on the corresponding plaintexts. We can,
therefore, carry out arithmetic operations such as addition,
subtraction, multiplication and division {}in the ciphertext field,
using the homomorphic properties of FHE
\cite{gentry2009fully,gentry2011implementing}. By restricting the
computation of SimRank similarity score in the ciphertext field, we
can achieve the protection of link information.

Further, we will show that PP-SimRank can keep the link information
remain hidden in a semi-honest security model.

\subsection{SimRank Overview}
\label{Sec:SimRank overview} SimRank \cite{jeh2002simrank} is a
similarity measure based on the linkage information in a network.
Two objects are regarded as similar ones if they are related to
similar objects. Specifically, let $S(v_i, v_j)$ denote the
similarity between two nodes $v_i$ and $v_j$ in a network $G = (V,
E)$. The iterative similarity computation equation of SimRank is as
follows:

{\arraycolsep=2pt
\begin{equation}
\begin{array}{rcl}
    S(v_i, v_j) & = & \displaystyle\frac{d_{out}}{|O(v_i)||O(v_j)|} %
    \sum_{k=1}^{|O(v_i)|}\sum_{l=1}^{|O(v_j)|} %
    S(O_{k}(v_i), O_{l}(v_j)),
\end{array}
\label{Eq:out SimRank}
\end{equation}
}

{\arraycolsep=2pt
\begin{equation}
\begin{array}{rcl}
    S(u_i, u_j) & = & \displaystyle\frac{d_{in}}{|I(u_i)||I(u_j)|} %
    \sum_{k=1}^{|I(u_i)|}\sum_{l=1}^{|I(u_j)|} %
    S(I_{k}(u_i), I_{l}(u_j)),
\end{array}
\label{Eq:in SimRank}
\end{equation}
}

\noindent where $O(v_i)$ or $O(v_j)$ denote the set of out-neighbors
of nodes $v_i$ or $v_j$, $I(u_i)$ or $I(u_j)$ denote the set of
in-neighbors of nodes $u_i$ or $u_j$, and $d_{out}$ $(d_{in})$ is
the decay factor.

\begin{figure}
    \centering
$\begin{array}{cc}
    \includegraphics[width=1.0in]{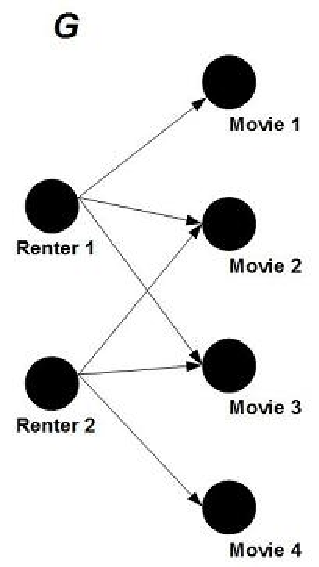} &
    \includegraphics[width=2.0in]{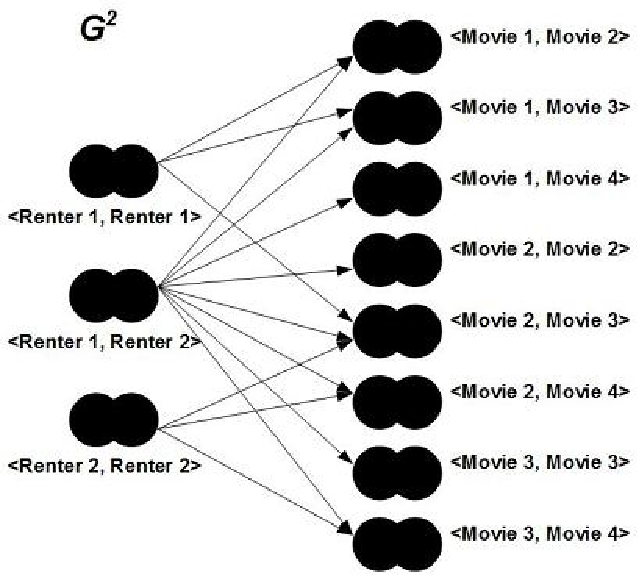} \\
(a) & (b) %
\end{array}$
    \caption{The combined renter data of online movie stores Alice and Bob.}
    \label{fig:simRankGMatrix}
\end{figure}

To simplify the SimRank score computation, SimRank derives a
node-pair graph {}$G^2 = (V^2, E^2)$. Each node $w$ in $V^2$ is
associated with a node pair $<a, b>$ in $G$, i.e., $a,b \in V$. The
presence of an edge between two nodes $<a, b>$ and $<c, d>$ in $G^2$
represents that there exist edges $(a, c)$ and $(b, d)$ in $G$ or,
there exist edges $(a, d)$ and $(b, c)$ in $G$. In $G^2$, each node
$w$ is associated with a SimRank score giving a measure of
similarity between the two nodes of G represented by $w$. The
neighbors of $w$ are the nodes whose similarity is needed to be
considered when the SimRank score of $w$ is computed.

Figure~\ref{fig:simRankGMatrix}(a) shows an example of the joint
rental data of two on-line movie stores Alice and Bob.
Figure~\ref{fig:simRankGMatrix}(b) is the corresponding node-pair
graph $G^2$, where the isolated nodes are omitted. There is an edge
in $G^2$ between the nodes $<Movie 1, Movie 2>$ and $<Renter 1,
Renter 2>$, because Renter 1 rents Movie 1 and Renter 2 rents Movie
2 in G.

In SimRank method, the node-pair graph $G^2$ is proposed to
explicitly figure out the propagation of similarity scores between
pairs of nodes. Later, we will further incorporate the node-pair
graph $G^2$ into the definition of similarity so that PP-SimRank is
able to calculate the SimRank score in ciphertext field.

\section{Privacy-preserving SimRank Protocol}
\label{Sec:3}

\begin{figure}
    \centering
    \includegraphics[width=3.2in]{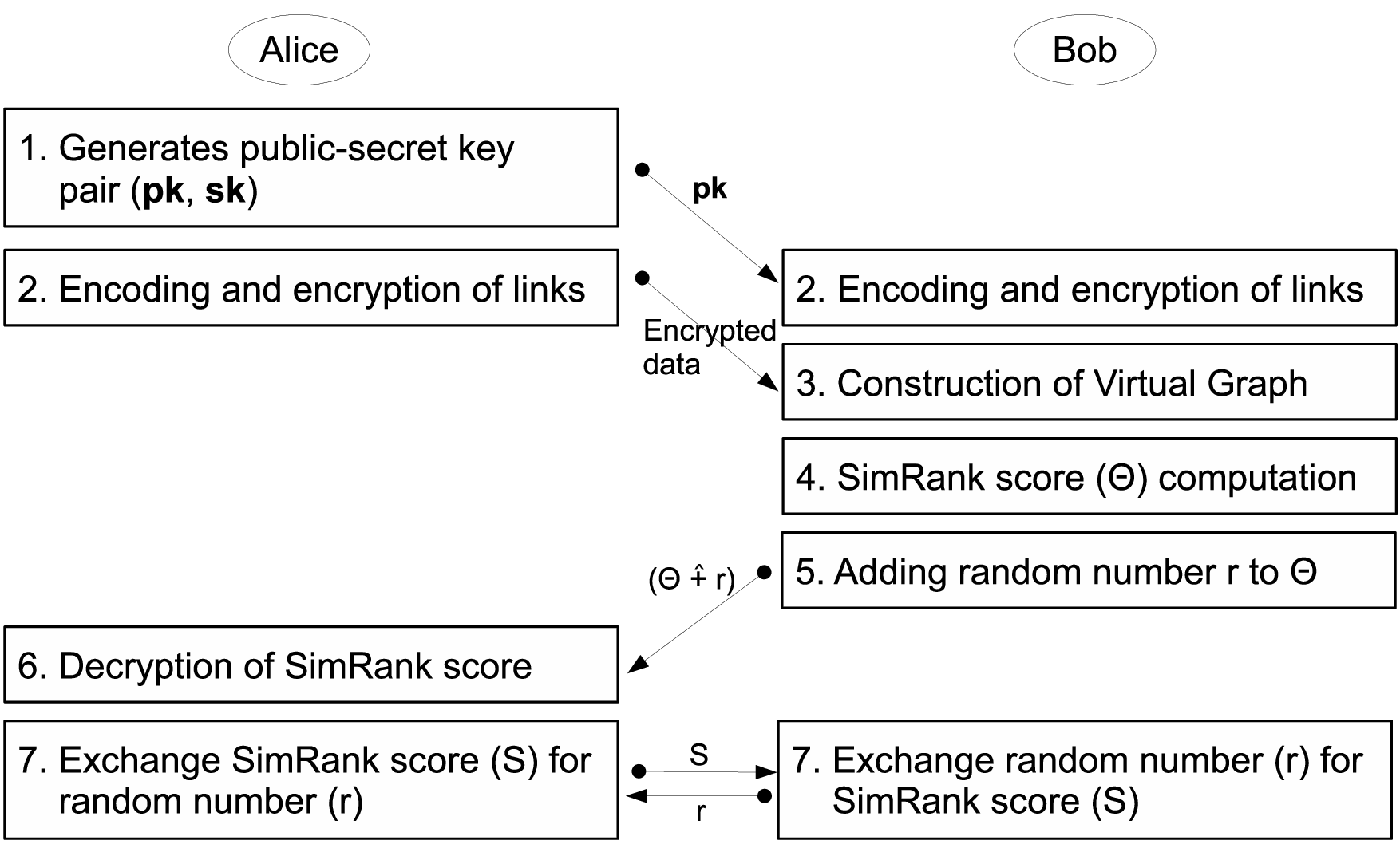}
    \caption{Privacy-preserving SimRank protocol.}
    \label{PP-SimRank protocol}
    \vspace{-6mm}
\end{figure}

\subsection{Protocol Overview}
We now introduce a new protocol called Privacy-Preserving SimRank
(PP-SimRank) based on fully-homomorphic encryption (FHE). In this
protocol, there are two essential roles: cryptographer and
calculator. Cryptographer determines the public-secret key pair
$(\textbf{pk}, \textbf{sk})$ used for encryption/decryption of data,
shares the public key with other while keeping the secret key
strictly to himself, and performs the decryption of encrypted
SimRank score. Calculator collects encrypted data from all parties,
constructs the virtual network in the ciphertext field, and performs
the SimRank score calculation on the virtual network. As
cryptographer does not have the virtual network and calculator does
not have the secret key for decryption, the SimRank score
computation over distributed information network can be achieved
while the link information of each parties' data is protected.

Figure~\ref{PP-SimRank protocol} shows the procedure of PP-SimRank
protocol. First, one party, called Alice, determines the
public-secret key pair $(\textbf{pk}, \textbf{sk})$, and sends the
public key $\textbf{pk}$ to the other party, called Bob. After that,
both Alice and Bob encode and encrypt their data with $\textbf{pk}$,
respectively, Alice then also transmits her encrypted data to Bob so
that Bob will hold a joint virtual information network $G = (V^A
\cup V^B, U, E^A \cup E^B)$ in the ciphertext field. Based on the
joint network $G$, Bob further builds a node-pair graph $G^2$.
Later, by incorporating the virtual node-pair network $G^2$ into the
definition of SimRank similarity, Bob performs the SimRank
calculation in ciphertext field. Finally, Bob adds a random number
to each of the final SimRank scores of the vertex pairs in $U^2$,
and sends the results to Alice for decryption. Both Alice and Bob
can then learn the SimRank similarity scores of the vertex pairs in
$U^2$ after they exchange the decrypted scores and the corresponding
added random numbers.

In this protocol, however, there are several challenging problems
needed to be solved. First, the bipartite network of each party has
to be encoded and encrypted in a way that allows the constructions
of the joint network $G$ as well as the corresponding node-pair
graph $G^2$ while the link protection is guaranteed. Second, based
on the node-pair graph $G^2$ built in ciphertext field, the whole
process of SimRank similarity computation has to be restricted in
the ciphertext field as well, since the FHE homomorphic properties
hold only if the plaintexts are encrypted with the same key.
Moreover, it makes the problem even more challenging that FHE has
the constraint of taking only 1 or 0 as inputs while the similarity
score is a real number in $[0, 1]$. We will tackle these challenges
step by step in the following subsections.

In Section \ref{Sec:EncodeEncrypt}, we explain how Alice and Bob
encrypt their data so that Bob is able to construct the graph matrix
of the node-pair graph $G^2$ in the ciphertext field. Section
\ref{Sec:ConstructionOfG2} shows how exactly the graph matrix of
$G^2$ is constructed. Based on $G^2$, Section
\ref{Sec:computationOfSimRankScore} gives the details of how the
SimRank score computation is achieved, when the computation is
restricted in the ciphertext field.

\subsection{Encoding and encryption of links}
\label{Sec:EncodeEncrypt}

We now explain how the parties encode and encrypt their data in
details. Without loss of generality, we assume that both Alice and
Bob know the total number of nodes in both $U$ and ($V^A \cup V^B$),
i.e., $m = |V^A|+|V^B|$ and $n = |U|$, and all the nodes are
ordered.

Note that the bipartite graph in our problem is not a weighted one.
The direct relationship between two nodes is either connection or
disconnection, i.e., a link $e \in \{0,1\}$. In many cryptographic
encryptions, it will output the same ciphertext given the same input
and encryption key{}. As a result, the link protection still cannot
be guaranteed even though the data is encrypted.

To tackle this challenge, we propose PP-SimRank based on FHE
\cite{gentry2009fully,gentry2011implementing}, which takes only 1 or
0 as possible input and outputs different ciphertexts for a given
input bit by introducing random noises in the encryption. That is,
after encryption, the ciphertext indicating the same bit varies. To
hide the link information more effectively, we let the parties
encode not only the connections but also the disconnections between
nodes into bit vectors and encrypt the bit vectors using FHE. For
Bob who receives Alice's data, he thus cannot distinguish the
ciphertexts corresponding to the 1- bits and 0- bits. The link
information is effectively protected. On the other hand, as the link
information is encoded into bits independently and encrypted without
other artificial noise (in addition to the random noise introduced
by FHE), Bob is able to utilize the homomorphic properties to
construct the virtual graphs $G$ and $G^2$.

Specifically, for each node $v_i$ in $V^A$ (or $V^B$), Alice (or
Bob) will generate a bit vector $E^A_{v_i}$ (or $E^B_{v_i}$), where
$E^A_{v_i}, E^B_{v_i} \in \{1, 0\}^n$, with respect to the
connections and disconnections of $v_i$ to all the nodes in $U$
based on her (his) own data. Each element $e_{i,j}$ in $E^A_{v_i}$
(or $E^B_{v_i}$) is $1$ if there is a link in $G^A$ (or $G^B$)
connecting $v_i$ and $u_j \in U$, or $0$ otherwise. After encoding
all the link information into bit vectors, the parties encrypt every
element $e_{i,j}$ in the bit vectors $E^A_{v_i}$ and $E^B_{v_i}$ to
get the cipher vectors $C^A_{v_i}$ and $C^B_{v_i}$, where each
element $c_{i,j} = Enc(e_{i,j})$, using the same public key
$\textbf{pk}$ (determined by Alice in the previous step). According
to the homomorphic properties of FHE, Bob can thus collect the
cipher vectors $C^A_{v_i}$ and $C^B_{v_i}$ to construct a virtual
joint bipartite graph in the ciphertext field.

\begin{figure}
    \centering
    \includegraphics[width=3in]{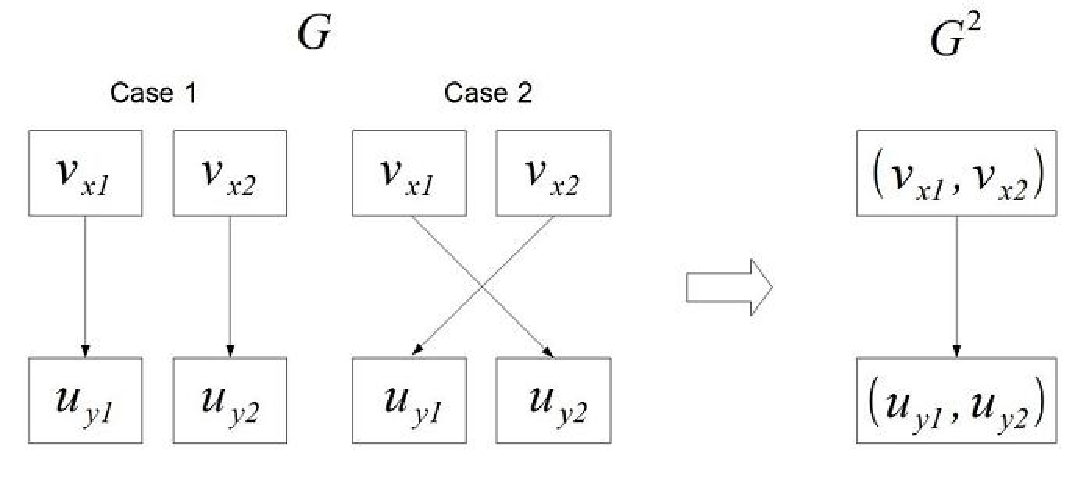}
    \caption{Two cases lead to an edge between $(v_{x1}, v_{x2})$ and $(u_{y1}, u_{y2})$.}
    \label{twoCases}
\end{figure}



\subsection {Construction of Virtual Graphs}
\label{Sec:ConstructionOfG2}

\begin{figure}
    \centering
    \includegraphics[width=3in]{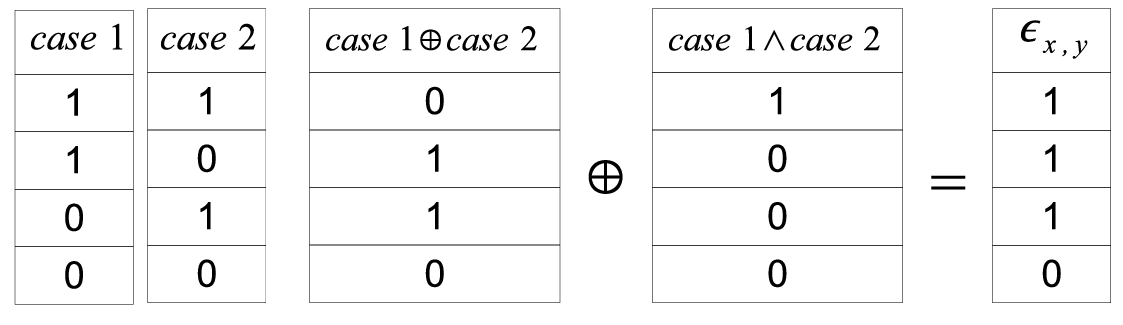}
    \caption{Or-equivalent operations.}
    \label{or-equivalent operation}
\end{figure}

This subsection shows how a joint virtual network $G^2$, where $G =
(V^A \cup V^B, U, E^A \cup E^B)$, is constructed in ciphertext
field.

To construct the virtual network $G^2$, knowing the link information
is essential. However, since the link information is hidden in the
cipher vectors $C^A_{v_i}$ and $C^B_{v_i}$, it is impossible for Bob
to construct a graph with explicit link information. Instead, we let
Bob construct the virtual graph $G^2$ by filling the graph matrix
$M$ of $G^2$ in ciphertext field. Each cipher element of $M$ can be
derived by applying the homomorphic operations on the cipher vectors
$C^A_{v_i}$ and $C^B_{v_i}$ as follows.

Specifically, let $w_x = <v_{x1}, v_{x2}>$ denotes a node-pair in
$G^2$, where $v$ is a node in ($V^A \cup V^B$), and $w_y = <u_{x1},
u_{x2}>$ denotes the node-pair of $U$ in $G^2$. Recall that, in
$G^2$, there is an edge $\epsilon_{x,y}$ connecting the nodes
$w_x=<v_{x1}, v_{x2}>$ and $w_y=<u_{y1}, u_{y2}>$ if and only if (as
shown in Figure~\ref{twoCases}), in $G$, at least one of the two
conditions in the following is satisfied.

\begin{enumerate}
\item $\{e_{x1,y1}$, $e_{x2,y2}\} \in (E^A \cup E^B)$: $u_{y1}$ is an out-neighbor of $v_{x1}$ and $u_{y2}$ is an out-neighbor of $v_{x2}$.
\item $\{e_{x1,y2}$, $e_{x2,y1}\} \in (E^A \cup E^B)$: $u_{y2}$ is an out-neighbor of $v_{x1}$ and $u_{y1}$ is an out-neighbor of $v_{x2}$.
\end{enumerate}

\noindent That is, the value of $\epsilon_{x,y}$ can be derived by
the following formula:

{\arraycolsep=2pt
\begin{equation}
\begin{array}{rcl}
    \epsilon_{x,y} & = & case1 \vee case2,
\end{array}
\label{Eq:or on plaintext 1}
\end{equation}
}

\noindent where

{\arraycolsep=2pt \setlength{\extrarowheight}{2mm}
\begin{equation}
\begin{array}{rcl}
\nonumber
    case1 = e_{x1,y1}\ \wedge\ e_{x2,y2},\\
    case2 = e_{x1,y2}\ \wedge\ e_{x2,y1}.
\end{array}
\end{equation}
}

\noindent When the OR-operation is carried out by XOR- and AND-
operations, Equation~(\ref{Eq:or on plaintext 1}) can be rewritten
as follows based on Figure~\ref{or-equivalent operation}.

{\arraycolsep=2pt
\begin{equation}
\begin{array}{rcl}
    \epsilon_{x,y} & = & (case 1 \oplus case 2) \oplus (case 1 \wedge case 2).
\end{array}
\label{Eq:or on plaintext 2}
\end{equation}
}

Now consider this problem in ciphertext field and let $\pi_{x, y}$
be the corresponding cipher of $\epsilon_{x, y}$ in $M$. Due to the
homomorphic properties of FHE, the XOR-/AND- operation on the
plaintext is equivalent to the addition/multiplication operation on
the corresponding ciphertext. The Value of $\pi_{x, y}$ can thus be
derived as follows.

{\arraycolsep=2pt \setlength{\extrarowheight}{2mm}
\begin{equation}
\begin{array}{rcl}
    \pi_{x,y} & = & Enc_{\textbf{pk}}(\epsilon_{x,y})\\

              & = & (Enc_{\textbf{pk}}(case 1) + Enc_{\textbf{pk}}(case 2))\\

              &   &  + (Enc_{\textbf{pk}}(case 1) \times Enc_{\textbf{pk}}(case 2)),\\
\end{array}
\label{Eq:or on ciphertext}
\end{equation}
} \vspace{-3mm} \noindent where \vspace{-3mm}

{\arraycolsep=2pt \setlength{\extrarowheight}{2mm}
\begin{equation}
\begin{array}{rcl}
\nonumber   Enc_{\textbf{pk}}(case 1) &=& c_{x1,y1} \times c_{x2,y2},\\

    Enc_{\textbf{pk}}(case 2) &=& c_{x1,y2} \times c_{x2,y1}.
\end{array}
\label{Eq:or on ciphertext 2}
\end{equation}
}

Consequently, Bob can construct a virtual graph $G^2$ for subsequent
SimRank score computation, while the link information is protected
and hidden in ciphers.

\subsection{Computation of SimRank score}
\label{Sec:computationOfSimRankScore}

In this subsection, we explain in details that how Bob computes the
SimRank score in ciphertext field, based on the virtual network
$G^2$. As Equations~(\ref{Eq:out SimRank}) and (\ref{Eq:in SimRank})
show, the SimRank similarity of two nodes $v_{x1}$ and $v_{x2}$ is
defined as the normalized sum of the similarities between their
neighbors. In the ciphertext field, however, the neighborhood
information as well as the link information is protected and hidden
in the encrypted graph matrix $M$ of $G^2$. To compute the SimRank
score, we then need to incorporate the graph matrix $M$ into
Equations~(\ref{Eq:out SimRank}) and (\ref{Eq:in SimRank}), and
express the formulas correspondingly in ciphertext field.

First, consider to incorporate the graph matrix $M$ into
Equations~(\ref{Eq:out SimRank}) and (\ref{Eq:in SimRank}). For $v
\in (V^A \cup V^B)$ and $u \in U$, let $w_x =<v_{x1}, v_{x2}>$ and
$w_y =<u_{y1}, u_{y2}>$ denote the node-pairs in $G^2$, and
$\epsilon_{x,y}$ be the element in $M$ that represents the
connection/disconnection between the node-pairs $w_x$ and $w_y$. The
SimRank score associating with $w_x$ and $w_y$ is:

{\arraycolsep=2pt \setlength{\extrarowheight}{3mm}
\begin{equation}
\begin{array}{lcr}
           S(w_x)&\\

            =  S(v_{x1}, v_{x2})&\\

            =  \displaystyle\frac{d_{out}}{|O(v_{x1})||O(v_{x2})|} %
           \sum_{i = 1}^{|O(v_{x1})|}\sum_{j = 1}^{|O(v_{x2})|} %
           S(O_i(v_{x1}),O_j(v_{x2}))&\\

            =  \displaystyle\frac{d_{out}}{\sum\limits_{i = 1}^ne_{x1, i}\sum\limits_{j = 1}^ne_{x2, j}} %
           \sum_{w_y \in G^2}S(w_y) \times \epsilon_{x, y},&
\end{array}
\label{Eq:w_x incorporate the graph matrix M }
\end{equation}
}

{\arraycolsep=2pt \setlength{\extrarowheight}{3mm}
\begin{equation}
\begin{array}{lcr}
           S(w_y)&\\

            =  S(u_{y1}, u_{y2})&\\

            =  \displaystyle\frac{d_{in}}{|I(u_{y1})||I(u_{y2})|} %
           \sum_{i = 1}^{|I(u_{y1})|}\sum_{j = 1}^{|I(u_{y2})|} %
           S(I_i(u_{y1}),I_j(u_{y2}))&\\

            =  \displaystyle\frac{d_{in}}{\sum\limits_{i = 1}^me_{i, y1}\sum\limits_{j = 1}^me_{j, y2}} %
           \sum_{w_x \in G^2}S(w_x) \times \epsilon_{x, y}.&
\end{array}
\label{Eq:w_y incorporate the graph matrix M}
\end{equation}
}

\noindent In Equation~(\ref{Eq:w_x incorporate the graph matrix M
}), the $\sum_{i = 1}^n e_{x1, i}$ ($\sum_{j = 1}^n e_{x2, j}$)
indicates the number of in-neighbors of $v_{x1}$ ($v_{x2}$), and
$S(w_y)$ is the SimRank similarity between $v_{x1}$'s potential
neighbor and $v_{x2}$'s potential neighbor. Whether a $S(w_y)$ score
will be counted in the $S(w_x)$ or not depends on the value of
$\epsilon_{x,y} \in \{0,1\}$ in the graph matrix $M$ of $G^2$.

Now we show the corresponding calculation of SimRank score in cipher
field. The challenge is that, both the network $G$ and the
corresponding node-pair graph $G^2$ are built in ciphers. We then
need to encrypt the SimRank similarity score using the same public
key $\textbf{pk}$, since the FHE homomorphic properties hold only
when the score is encrypted in the same field as the link
information.

Specifically, note that the FHE method takes only binary integer,
i.e., $\{1/0\}$, as inputs, while the similarity score $S(w_x)$ is a
real number $\in [0,1]$. To allow the encryption of $S(w_x)$ using
FHE, we encode the real number $S(w_x) \in [0,1]$ (as well as
$d_{in}$ and $d_{out}$) into binary representation, i.e., a bit
string, and encrypt the bit string into cipher string. The
arithmetic operations on the real numbers then change to the binary
arithmetic operations on the (bit/cipher) string. Let
$\hat{\times}$, $\hat{\div}$, and $\hat{\sum}$ denote the
multiplication, division and sum binary arithmetic operations,
respectively. We discuss the implementations of the binary
arithmetic operations in Section~\ref{Sec:ArithmeticOperations}, and
convert Equations~(\ref{Eq:w_x incorporate the graph matrix M }) and
(\ref{Eq:w_y incorporate the graph matrix M}) into the following
formulas in ciphertext field.

{\arraycolsep=2pt \setlength{\extrarowheight}{3mm}
\begin{equation}
\begin{array}{rcl}
    \Theta(w_x) & = & Enc_{\textbf{pk}}(S(w_x))\\

                & = & \Delta \ \displaystyle\hat{\times}\ \hat{\sum}_{w_y \in G^2} (\Theta(w_y)\ \hat{\times}\ \pi_{x, y}),\\
\end{array}
\label{Eq:w_x SimRank in ciphertext field 1}
\end{equation}

\setlength{\extrarowheight}{3mm}
\begin{equation}
\begin{array}{rcl}
    \Theta(w_y) & = & Enc_{\textbf{pk}}(S(w_y))\\

                & = & \Square \ \displaystyle\hat{\times}\ \hat{\sum}_{w_x \in G^2} (\Theta(w_x)\ \hat{\times}\ \pi_{x, y}),\\
\end{array}
\label{Eq:w_y SimRank in ciphertext field 1}
\end{equation}

\noindent where

\begin{equation}
\begin{array}{rcl}
\nonumber
    \Delta & = & \displaystyle Enc_{\textbf{pk}}(d_{out})\ \hat{\div}\ (\hat{\sum}_{i = 1}^n c_{x1, i}\ \hat{\times}\ \hat{\sum}_{j = 1}^n c_{x2, j}),
\end{array}
\label{Eq:w_x SimRank in ciphertext field 2}
\end{equation}

\vspace{-3mm}

\begin{equation}
\begin{array}{rcl}
\nonumber
    \Square & = & \displaystyle Enc_{\textbf{pk}}(d_{in})\ \hat{\div}\ (\hat{\sum}_{i = 1}^m c_{i, y1}\ \hat{\times}\ \hat{\sum}_{j = 1}^m c_{j, y2}).
\end{array}
\label{Eq:w_y SimRank in ciphertext field 2}
\end{equation}
}

According to Equations~(\ref{Eq:or on ciphertext}), (\ref{Eq:w_x
SimRank in ciphertext field 1}) and (\ref{Eq:w_y SimRank in
ciphertext field 1}), Bob is able to perform the whole process of
SimRank similarity calculation in ciphers without knowing the
explicit link information of Alice's data. PP-SimRank, therefore,
achieves a secure similarity measure on a joint networks consisting
of different parties' data.

\section{Implementation Issues} \label{ch:s5}
\label{Sec:4}

\subsection{Arithmetic operations in the ciphertext field}
\label{Sec:ArithmeticOperations}

As mentioned in Section~\ref{Sec:computationOfSimRankScore}, we
implement the binary arithmetic operations to perform the addition,
multiplication and division on the cipher strings representing
$\Theta(w)$. First, we implement two virtual circuit functions, a
full-adder function and a full-subtractor function, that can
simulate the functionality of a real full-adder and full-subtractor
circuit using the fully-homomorphic properties of FHE, respectively.
By combining a sequence of the full-adder (or full-subtractor)
function, we carry out a binary addition $\hat{+}$ (or subtraction
$\hat{-}$) function that can perform the addition (or subtraction)
of two cipher strings. After that, the multiplication function
$\hat{\times}$ can be developed by calculating the partial products
of the input cipher strings, shifting each of the product to the
left, and adding them together using the addition function
$\hat{+}$. Similarity, division function $\hat{\div}$ consists of
subtraction function $\hat{-}$ and the shifting mechanism. We
explain the details in the following.

\begin{figure}[t]
\centering $\begin{array}{c}
    \includegraphics[width=2.5in]{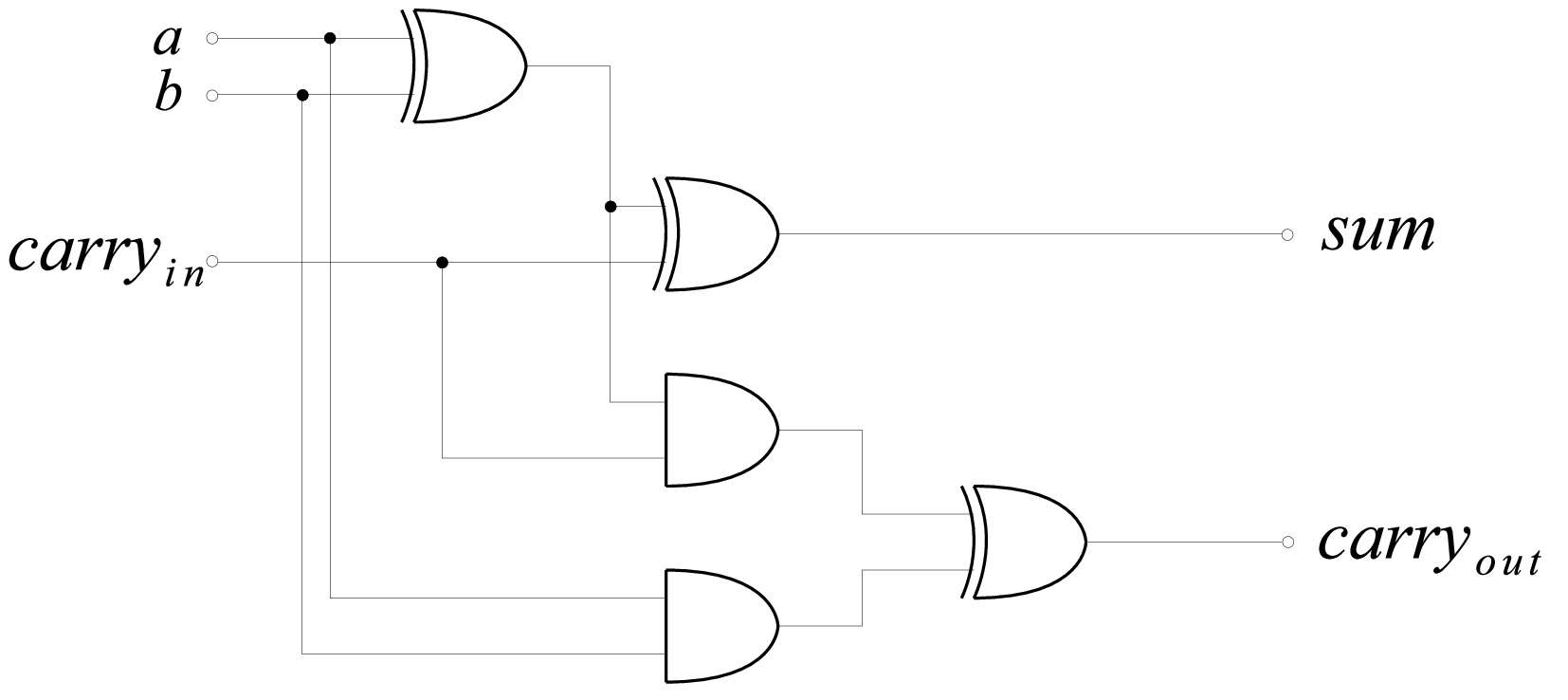}\\
    (a)\\

    \includegraphics[width=2.5in]{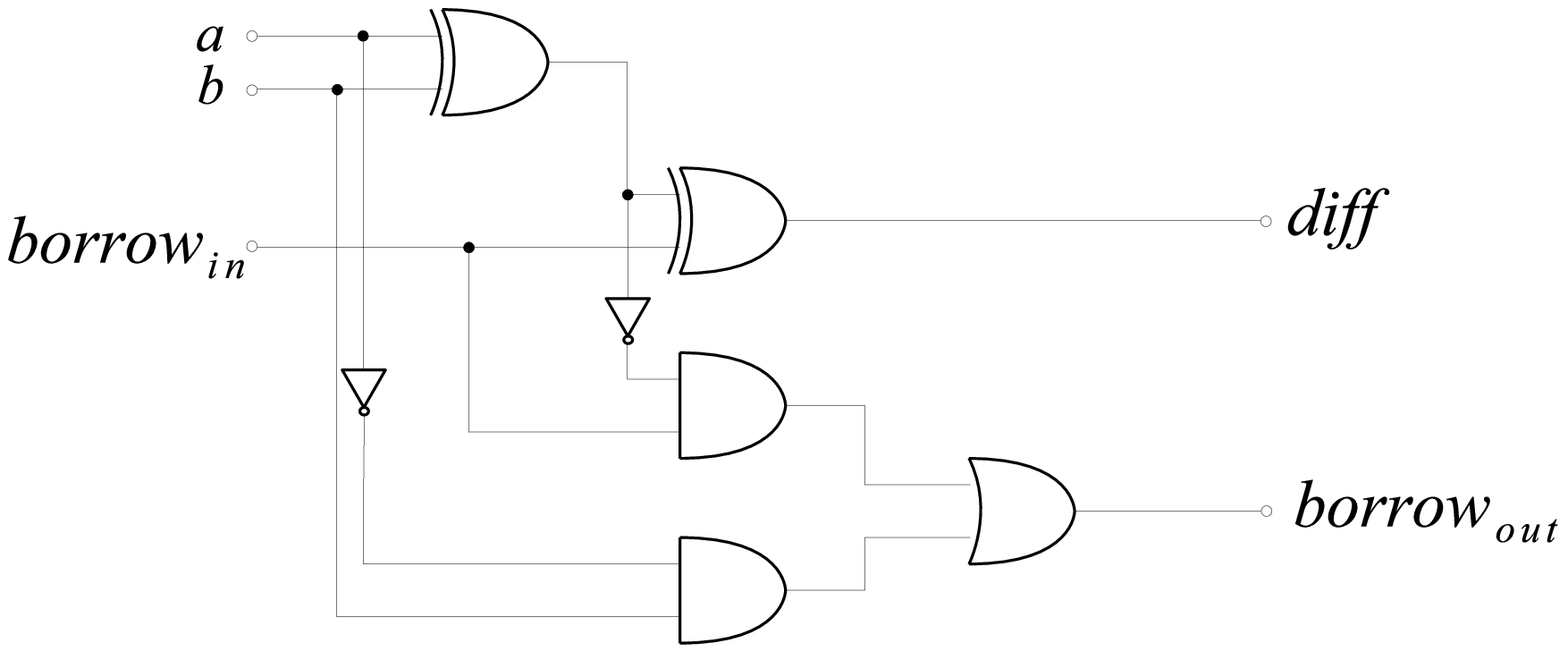}   \\
    (b)
\end{array}$
\caption{Logical circuits of (a) a full-adder and (b) a
full-subtractor.}
    \label{full_adder_subtractor}
\end{figure}


\textbf{Binary addition $\hat{+}$.} As shown in
Figure~(\ref{full_adder_subtractor})(a), in electronics, a real
full-adder is composed by a number of logical operations. The real
full-adder adds the operand bits $a$ and $b$ with the carried-in bit
$carry_{in}$ received from the previous full-adder, and output the
answer bit $sum$ and carry-out bit $carry_{out}$ as the result. The
operations of a real full-adder can be expressed in terms of XOR and
AND as follows:

{\arraycolsep=2pt \setlength{\extrarowheight}{2mm}
\begin{equation}
\begin{array}{rcl}
    sum & = & a\ \oplus\ b\ \oplus\ carry_{in},\\

    carry_{out} & = & (a \wedge b)\ \oplus\ (carry_{in}\ \wedge\ (a \oplus b)).\\
\end{array}
\label{Eq:full adder on plaintext}
\end{equation}
}

\noindent As these operations are in bit level the same as the
inputs of FHE, we then can convert the real full-adder into
ciphertext field using the fully-homomorphic properties. By
replacing the XOR- and AND- operations with addition and
multiplication, respectively, Equation~(\ref{Eq:full adder on
plaintext}) is transformed into the following formulas:

{\arraycolsep=2pt \setlength{\extrarowheight}{2mm}
\begin{equation}
\begin{array}{rcl}
    c_{sum} & = & c_a + c_b + c_{carry_{in}},\\

    c_{carry_{out}} & = & (c_a \times c_b) + (c_{carry_{in}} \times (c_a + c_b)),\\
\end{array}
\label{Eq:full adder on ciphertext}
\end{equation}
}

\noindent where $c_{sum}$, $c_a$, $c_b$ and $c_{carry}$ denote the
corresponding ciphers of $sum$, $a$, $b$ and $carry$, respectively.
Based on Equation~(\ref{Eq:full adder on ciphertext}), we then
develop a function, called virtual full-adder function which
executes the functionality of the real full-adder circuit in
ciphertext field. After that, let each similarity score $S(w)$ in
PP-SimRank use $l$ bits for its binary representation. By connecting
$l$ virtual full-adder function together, i.e., each function takes
the output cipher $c_{carry_{out}}$ of previous full-adder function
as its input carry-in cipher $c_{carry_{in}}$, we can carry out a
binary addition function $\hat{+}$ to add cipher strings of
$\Theta(w)$ in ciphertext field.

\textbf{Binary multiplication $\hat{\times}$.} We further utilize
the binary addition function $\hat{+}$ to implement the binary
multiplication function $\hat{\times}$. Specifically, similar to
what we usually do in the computation of multiplying two decimal
numbers, the binary multiplication function $\hat{\times}$ can be
carried out by computing a sequence of partial products, shifting
the resulting partial products to the left, and finally applying the
binary addition fuction $\hat{+}$ to add the products together.

Similarly, we begin from the real full-subtractor circuit and then
explain the implementations of binary subtraction $\hat{-}$ and
binary devision $\hat{\div}$ operations in ciphertext field.

\textbf{Binary subtraction $\hat{-}$.} The logical operations of a
real full-subtractor shown in Figure~\ref{full_adder_subtractor}(b)
can be expressed as follows:

{\arraycolsep=2pt \setlength{\extrarowheight}{2mm}
\begin{equation}
\begin{array}{rcl}
    diff & = & a \oplus b \oplus borrow_{in},\\

    borrow_{out} & = & (\ \overline{a}\ \wedge b)\ \vee\ (borrow_{in}\ \wedge\ \overline{(a \oplus b)}\ )\\

    & = & ((1 \oplus a) \wedge b) \vee (borrow_{in} \wedge (1 \oplus (a \oplus b))),
\end{array}
\label{Eq:full subtractor on plaintext}
\end{equation}
}

\noindent where the NOT-operation on a bit $a$ is achieved by
computing the value of ($1\ \oplus\ a$). Let

{\arraycolsep=2pt \setlength{\extrarowheight}{2mm}
\begin{equation}
\begin{array}{rcl}
    t1 & = &  (1\ \oplus\ a)\ \wedge\ b,\\

    t2 & = & borrow_{in}\ \wedge\ (1\ \oplus\ (a\ \oplus\ b).
\end{array}
\label{Eq:full subtractor on plaintext t1 t2}
\end{equation}
}

\noindent Based on the truth tables in Figure~\ref{or-equivalent
operation}, $borrow_{out}$ can also be expressed in terms of XOR-
and AND- logics.

{\arraycolsep=2pt
\begin{equation}
\begin{array}{rcl}
    borrow_{out} & = & (t1\ \oplus\ t2)\ \oplus\ (t1\ \wedge\ t2).
\end{array}
\label{Eq:full subtractor on plaintext b}
\end{equation}
}

\noindent Accordingly, a virtual full-subtractor in ciphertext field
is derived in Equation~(\ref{Eq:full subtractor on ciphertext 1})
(from Equations~(\ref{Eq:full subtractor on plaintext}),
(\ref{Eq:full subtractor on plaintext t1 t2}) and (\ref{Eq:full
subtractor on plaintext b})) again by substituting the XOR- and AND-
logics with addition and multiplication, respectively.

{\arraycolsep=2pt \setlength{\extrarowheight}{2mm}
\begin{equation}
\begin{array}{rcl}
    c_{diff} & = & c_a + c_b + c_{borrow_{in}},\\

    c_{borrow_{out}} & = & (c_{t1} + c_{t2}) + (c_{t1} \times c_{t2}),
\end{array}
\label{Eq:full subtractor on ciphertext 1}
\end{equation}
}

\noindent where

{\arraycolsep=2pt \setlength{\extrarowheight}{2mm}
\begin{equation}
\begin{array}{rcl}
\nonumber
    c_{t1} & = & (c_1 + c_a) \times c_b,\\

    c_{t2} & = & (c_{borrow_{in}} \times (c_1 + (c_a + c_b)).
\end{array}
\label{Eq:full subtractor on ciphertext 2}
\end{equation}
}

\noindent By regarding a virtual full-subtractor as a basic unit, we
can thus implement a binary subtractor $\hat{-}$ in a similar way of
building a binary addition $\hat{+}$, where each function takes the
output cipher $c_{borrow_{out}}$ of previous full-subtractor
function as its input borrow-in cipher $c_{borrow_{in}}$.

\textbf{Binary division $\hat{\div}$.} The binary division function
$\hat{\div}$ also can be carried out by utilizing a sequence of
binary subtraction function $\hat{-}$. Similar to the decimal
division, the binary division function $\hat{\div}$ computes the
quotient cipher string by iteratively subtracting divisor from the
dividend using binary subtraction function $\hat{-}$ and shifting
the divisor to the right after each binary subtraction $\hat{-}$.
Each ciphertext in the quotient cipher string is determined by
adding the output borrow cipher $c_{borrow_{out}}$ of the binary
subtraction $\hat{-}$ with a cipher $c_1 = Enc_{\textbf{pk}}(1)$,
which is equivalent to the NOT-operation ($1 \oplus borrow_{out}$)
in the plaintext field.

\subsection{Efficiency improvement}
\label{Sec:Efficiency improvement}

FHE encrypts bits into ciphertexts (with random noises) in the form
of very big integers to achieve semantic security. A side effect,
however, is that operations on big integers usually cost a lot of
time. In this subsection, we propose to reduce the execution time of
PP-SimRank in three ways: reducing the number of binary
multiplications in the score calculations, replacing the FHE
encryption with  look-up, and carrying out the FHE re-encryption
mechanism by Re-encryption protocol instead.

\vspace{1mm}

\textbf{Reducing the number of binary multiplications.}
Equations~(\ref{Eq:w_x SimRank in ciphertext field 1}) and
(\ref{Eq:w_y SimRank in ciphertext field 1}) give the detailed
calculations of SimRank similarity in ciphertext field. We note that
the execution efficiency can be improved by replacing the binary
multiplication $\hat{\times}$ of ($\Theta(w_x)\ \hat{\times}\ \pi$)
with simple multiplications $\times$. Specifically, recall that as
mentioned in Section~\ref{Sec:ConstructionOfG2}, a ciphertext $\pi$
represents a bit $\epsilon$ in the plaintext, and each SimRank score
$\Theta(w_x)$ in the ciphertext field is represented in the form of
a cipher string corresponding to the binary representation (a bit
string) of $S(w_x)$ in plaintexts. Because $\epsilon$ is either $0$
or $1$, multiplying $\epsilon$ with the bit sting of $S(w_x)$ can be
carried out by AND- operations between them. Correspondingly, in the
ciphertext field, we can achieve ($\Theta(w_x)\ \hat{\times}\ \pi$)
in Equations~(\ref{Eq:w_x SimRank in ciphertext field 1}) and
(\ref{Eq:w_y SimRank in ciphertext field 1}) by multiplying
(corresponding to AND-operation in plaintext field) $\pi$ with every
ciphertext in the cipher string of $\Theta(w_x)$, instead of
performing the time consuming binary multiplication.

\begin{table}[t]
\centering \caption{The execution time of FHE encryption, FHE
re-encryption and Ciphertable look-up w.r.t. various security
levels.} \label{Table:Encryption Time}
\begin{tabular}{| c | c | c | c |}
  \hline
  & FHE & FHE &   CipherTable \\
  Security Level & Encryption & Re-Encryption & Look-up\\
  (dimension) & (sec) & (sec) & (sec)\\
  \hline
  512 & 0.217  & 6.3 & 0.020 \\
  \hline
  2048 & 2.084  & 34.1 & 0.254 \\
  \hline
  8192 & 20.300 & 185 & 2.397 \\
  \hline
\end{tabular}
\end{table}

\vspace{1mm}

\textbf{CipherTable look-up.} Table \ref{Table:Encryption Time}
shows the FHE encryption time of one bit with respect to different
security levels\footnote{In \cite{gentry2011implementing}, they use
lattices of 512, 2048 and 8192 dimensions to provide security levels
called as "toy", "small" and
"medium".\label{Footnote:1}}. Accordingly, it would be
infeasible if Alice and Bob encrypt their data by calling the FHE
encryption function.

To solve this problem, we use CipherTable look-up instead which is
at least 8 times faster than FHE encryption according to the results
in Table \ref{Table:Encryption Time}. That is, since there are only
two possible inputs $1$ and $0$ of FHE, Alice and Bob can share the
public key $\textbf{pk}$ in advance and respectively prepare a
CipherTable containing a set of ciphertexts w.r.t. the plaintexts
$\{0, 1\}$ before they really start computing the similarity
measures on their joint data. Therefore, when PP-SimRank is
performed, Alice and Bob can encrypt their data in step 2 of
Figure~\ref{PP-SimRank protocol} by simply choosing a corresponding
ciphertext from their own CipherTable, instead of calling FHE
encryption function. As FHE outputs different ciphertexts (by
introducing a random noise in the encryption) and the CipherTable is
built by each party itself, this replacement is satisfactory.

\vspace{1mm}

\textbf{Re-encryption protocol.} The PP-SimRank protocol protects
the link information using FHE. For a given bit representing direct
connection or disconnection, FHE will introduce a random noise into
the encryption to output different ciphertexts. Due to the random
noise, the inverse mapping from ciphers to the bit becomes
challenging. However, the noise in a ciphertext will propagate and
grow with the homomorphic operations operated on it. When the
magnitude of noise reaches a certain limitation\footnote{Given the
public key, FHE \cite{gentry2011implementing} allows one to test the
magnitude of noise in a ciphertext. If the noise grows beyond a
certain ratio of the decryption radius, there is a demand of
re-encryption. \label{footnote test noise}}, the ciphertext will not
be able to be correctly decrypted. A re-encryption mechanism that
can refresh the ciphertext by reducing the associated noise is thus
required.

The work in \cite{gentry2011implementing} has suggested a mechanism,
called bootstrapping, for re-encryption. The idea of bootstrapping
is to refresh a ciphertext by decrypting the ciphertext
homomorphically with the encrypted secret key. As shown in Table
\ref{Table:Encryption Time}, however, the re-encryptions of
bootstrapping under various security level are very time-consuming.
Instead, we propose an more efficient re-encryption protocol for our
problem scenario. The results in Table \ref{Table:Encryption Time}
shows that our re-encryption protocol runs at least $77$ times
faster than FHE re-encryption method.

Specifically, when it is detected\footnote{Refer to Footnote
\ref{footnote test noise}.}, by Bob, that the magnitude of noise
associating with a ciphertext reaches a limitation during the
computation, Bob asks Alice's help for re-encryption. The noisy
ciphertext is sent to Alice. Alice then decrypts the noisy
ciphertext using the secret key $\textbf{sk}$ (which is kept
strictly to Alice), and encrypts the resulting bit again using the
same public key $\textbf{pk}$. Here, as mentioned previously, Alice
can perform the encryption by looking up her CipherTable, rather
than call the FHE encryption function. The re-encryption efficiency
is thus improved significantly. After that, a corresponding new
ciphertext is sent back to Bob for subsequent computation.

This re-encryption protocol will not damage the private link
protection. For Bob, what he sends out and receives are both
ciphertexts. Therefore, he cannot get additional information about
the links in Alice' data. On the other hand, although Alice knows
the true value of the received bit by decryption, she does not know
the semantic meaning of the bit since a bit can be any link
information in Bob's data, any bit in the binary representation of a
similarity score between any two vertices, or even a temporary bit
appearing only during the computation process. Consequently, the
re-encryption protocol is efficient and satisfactory for our problem
scenario.

\section{Performance analysis} \label{ch:s5}
\label{Sec:5}

\subsection{Security of PP-SimRank}
The purpose of PP-SimRank is to achieve SimRank similarity score
co-computation based on a joint virtual network $G = (V^A \cup V^B,
U, E^A \cup E^B)$ while protecting every link information in both
$E^A$ and $E^B$ from being stolen. In the analysis, we ask two
questions: (1) Can Bob who has the whole network (in virtual) derive
the links in Alice's data by generating many plaintext-ciphertext
pairs with the given public key \textbf{pk}? (2) Can Alice or Bob
derive the links in the other's data after the similarities between
nodes in $U$ is revealed?

The first question is also known as a birthday attack
\cite{kemeny1966introduction}. Since, in PP-SimRank, the public key
\textbf{pk} generated by Alice is shared with Bob, Bob is able to
generate cipher-plaintext pairs by himself. The cipher-plaintext
pairs he generated may contain a cipher that collides with an
encrypted data received from Alice. The link information in $E^A$,
therefore, has a chance to be stolen by Bob. However, according to
the study \cite{gentry2011implementing}, the ciphertext space of FHE
is enormous large, i.e., ciphertext $c \in \mathbb Z_d = \{1, 2,
\dots, d\}$ where $\log_2 d \in \{195764, 785006, 3148249\}$ w.r.t.
the security level toy, small and median. Under the analysis of the
birthday problem \cite{kemeny1966introduction}, the expected number
of ciphertexts Bob has to generate before he finds the first
collision is about $\sqrt{\dfrac{\pi}{2}d}$, which is still
significantly large. Hence, the probability that Bob finds a
collision and steals the link information from Alice is negligibly
small.

For the second question, the answer is ``No''. The reason is that,
in addition to having the pairwise similarity scores $S(w_y)$
between nodes in $U$, a party also needs the pairwise similarity
scores $S(w_x)$ between the nodes in $(V^A \cup V^B)$ so that the
link information of the other party can be derived by solving linear
algebra problems. However, at the end of PP-SimRank protocol, the
similarity score $S(w_x)$ between the nodes in $(V^A \cup V^B)$ are
not available for both Alice and Bob since Alice does not have the
data and Bob does not have the secret key for decryption.
Consequently, in a semi-honest security model, PP-SimRank does not
give a chance to a party to steal the link information of the
others.

\subsection{Communication and computation complexity}

We illustrate the communication and computational cost of PP-SimRank
on a joint network $G=(V^A \cup V^B, U, E^A \cup E^B)$ in this
section. Assume there are $m$ nodes in $(V^A \cup V^B)$, $n$ nodes
in $U$, and the dimension of the node-pair graph matrix $M$ is
$C^m_2 \times C^n_2$, i.e., $O(m^2) \times O(n^2)$. Let $l$ be the
number of bits used for binary representation of a SimRank score in
plaintext, and $k$ be the number of bits for a ciphertext
transmission. First, we analyze the computational cost on the
calculator, Bob. To construct the virtual node-pair graph $G^2$ in
step 3 of PP-SimRank, Bob computes each element $\pi_{x, y}$ in $M$
based on Equations~(\ref{Eq:or on ciphertext}), consisting of two
additions $+$ and three multiplications $\times$. Therefore, Bob
performs totally $O(m^2n^2)$ additions and multiplications to fill
the whole graph matrix. Next, in step 4 of PP-SimRank, Bob performs
the SimRank score computation in ciphertext field according to
Equations~(\ref{Eq:w_x SimRank in ciphertext field 1}) and
(\ref{Eq:w_y SimRank in ciphertext field 1}). For each score
$\Theta(w_x)$ (or $\Theta(w_y)$), the required arithmetic operations
on the cipher string{} include $O(1)$ binary division $\hat{\div}$,
$O(n^2)$ (or $O(m^2)$) binary additions $\hat{+}$ and $O(n^2 + 1)$
(or $O(m^2 + 1)$) binary multiplications $\hat{\times}$, because the
number of nodes in $(V^A \cup V^B)^2$ and $U^2$ are $O(m^2)$ and
$O(n^2)$, respectively. We can further improve the performance by
replacing the  binary multiplication of ($\Theta(w)\ \hat{\times}\
\pi$) in Equations~(\ref{Eq:w_x SimRank in ciphertext field 1}) and
(\ref{Eq:w_y SimRank in ciphertext field 1}) with $l$ simple
multiplications $\times$ as explained in Section~\ref{Sec:Efficiency
improvement}. The complexity for computing a $\Theta(w_x)$ (or
$\Theta(w_y)$) therefore becomes $O(1)$ binary division
$\hat{\div}$, $O(n^2)$ (or $O(m^2)$) binary additions $\hat{+}$,
$O(n^2l)$ (or $O(m^2l)$) simple multiplications and $O(1)$ binary
multiplication. The total binary arithmetic operations required in
one iteration to compute all the SimRank scores are thus $O(m^2 +
n^2)$ binary divisions, $O(m^2 + n^2)$ binary multiplications,
$O(m^2n^2)$ binary additions and $O(m^2n^2l)$ simple
multiplications. Table \ref{Table:binary operation running time}
shows the execution efficiency of each (binary) arithmetic operation
in ciphertext field.

\begin{table}
\begin{center}
\caption{The execution time (sec) of (binary) arithmetic operations
in ciphertext field.} \label{Table:binary operation running time}
\begin{tabular}{ c | c | c | c |}
\hline
\multicolumn{1}{|c|}{\multirow{2}{*}{Arithmetic Operation}}&\multicolumn{3}{c|}{Security Level (dimension)}\\
\cline{2-4}
\multicolumn{1}{|c|}{}& 512 & 2048 & 8192 \\
\hline
\multicolumn{1}{|c|}{{${+}$}} & $1.0 \times 10^{-5}$ & $3.0 \times 10^{-5}$ & $1.4 \times 10^{-4}$\\
\hline
\multicolumn{1}{|c|}{$\times$} &$4.67 \times 10^{-3}$ & 0.03& 0.135\\
\hline
\multicolumn{1}{|c|}{$\hat{+}$} & 0.81 & 5.58&54.92\\
\hline
\multicolumn{1}{|c|}{$\hat{-}$} &1.11 & 8.61 & 92.7\\
\hline
\multicolumn{1}{|c|}{$\hat{\times}$} & 5.30 & {47.06} & 584.00 \\
\hline
\multicolumn{1}{|c|}{$\hat{\div}$} & 22.39 & 191.94 & 1961.00\\
\hline
\end{tabular}
\end{center}
\end{table}

\begin{table}
\begin{center}
\caption{The numbers of re-encryptions involved during various
operations.} \label{Table:re-encrypt counts}
\setlength{\extrarowheight}{4pt}
\begin{tabular}{ c | c | c | c |}
\hline
\multicolumn{1}{|c|}{\multirow{2}{*}{Arithmetic Operation}}&\multicolumn{3}{c|}{Num. of Re-encryptions}\\
\cline{2-4}
\multicolumn{1}{|c|}{}& max & Avg. & min \\
\hline
\multicolumn{1}{|c|}{{$+, \times, \hat{+}, \hat{-}$}} & 0 & 0 & 0\\
\hline
\multicolumn{1}{|c|}{$\hat{\times}$} & 40 & 36.04 & 33 \\
\hline
\multicolumn{1}{|c|}{$\hat{\div}$} & 280 & 224.31 & 191\\
\hline
\end{tabular}
\end{center}
\end{table}

In PP-SimRank, the communication between Alice and Bob occurs in
three steps, including encrypted link data transmission after step
$2$, resulting cipher SimRank score transmission after step $5$ and
re-encrypt protocol required during computation. First, assuming
that Alice has $|V^A|$ nodes that exclusively owned by herself. She
thus needs totally {}($n|V^A|$) bits to encode all the link
information between $V^A$ and $U$ and transmits totally $O(kn|V^A|)$
bits data to Bob for subsequent calculations. Second, after
completing the SimRank score computation, Bob will send $O(n^2)$
noisy cipher SimRank scores $\Theta'(w_y) = \Theta(w_y) \hat{+} r_y$
to Alice after step $5$. Since each cipher score is represented by
$l$ ciphers and each cipher needs $k$ bits for transmission, the
communication cost is $O(n^2lk)$ bits. Finally, as shown in Table
\ref{Table:re-encrypt counts}, the times of re-encryption required
vary with binary arithmetic operations on different ciphertexts.
Here we evaluate the communication cost in one re-encryption
protocol: Alice and Bob send one cipher to each other. Therefore,
the cost is $O(k)$ bits.

\section{Extension to multi-party scenario}
\label{Sec:6} In this section, we further extend the PP-SimRank
protocol to a more general case, i.e., multi-party scenario. We
assume there are totally $s$ parties and the $i$-th party holds an
information network $G^i=(V^i, U, E^i)$ where the vertex sets $V^1$,
$V^2$, $\dots $, and $V^s$ are mutually disjoint. The purpose is to
let all parties learn the similarity between nodes in $U$ based on
their joint network $G=(\cup V^i, U, \cup  E^i)$ while the links
hold by each party are strictly known to itself.

Similar to the two-party scenario, here PP-SimRank also asks two of
the $s$ parties to act as a cryptographer (Alice) and a calculator
(Bob) and to perform their duty respectively. For the rest parties,
they encrypt their data with the public key \textbf{pk} received
from the cryptographer, and then transmit the encrypted data to the
calculator after step 2 of PP-SimRank protocol (shown in
Figure~(\ref{PP-SimRank protocol})). In addition, each of the rest
parties also determines a random number list $R^i$ containing
$C^n_2$ elements, encrypts each element $r^i_y$, $y = \{1, \dots,
C^n_2\}$, with \textbf{pk}, and sends the encrypted random numbers
to the calculator. After the calculation of SimRank similarity is
completed in step 4, the calculator adds each of the resulting
SimRank scores $\Theta (w_y)$ with a set of random numbers {$r^1_y,
r^2_y, \dots, r^{s-1}_y$} collected from all parties except the
cryptographer. Therefore, at the end, the cryptographer will hold
the noisy scores $S'(w_y)$ after decryption, and each of the other
parties will have a list of random numbers that have been added to
the noisy scores $S'(w_y)$. The correct SimRank scores then can be
derived after all parties exchange these information with each
other.

Now consider the corresponding complexities. Assume there are
totally $m$ nodes in $(V^1 \cup V^2 \cup \dots \cup V^s)$, i.e., $m
= |V^1|+|V^2|+\dots+|V^s|)$ and $n$ nodes in $U$, i.e., $n=|U|$. The
dimension of the node-pair graph matrix $M$ is therefore $O(m^2)
\times\ O(n^2)$, which is the same with the dimension in the
two-party scenario. As the result, the computational complexity of
multi-party PP-SimRank is identical to that we have analyzed in the
two-party scenario. On the other hand, the communication cost
slightly increases. Assume that each score uses $l$ bits for its
binary representation and each ciphertext needs $k$ bits for its
transmission. There is a new cost $O((s-2)n^2lk)$ for transmitting
$(s-2)$ encrypted lists of random numbers from all parties, except
the cryptographer and the calculator, to the calculator, while the
cost for encrypted link data transmission becomes $O((m - |V^B|)nk)$
bits. The cost for transmitting the resulting cipher SimRank scores
remains the same.

\section{Conclusion}
\label{Sec:7}

In this paper, we have addressed the problem of privacy-preserving
co-computation of link-based similarity measure in a distributed
bipartite network, and proposed a new privacy-preserving protocol,
PP-SimRank, as a solution. PP-SimRank strictly protects each party's
link information and reveals nothing but the desired similarity
measures on the (virtual) joint network. We also implemented the
basic binary arithmetic operations in ciphertext field to securely
compute SimRank scores under fully-homomorphic encryption (FHE) and
demonstrated the potential of FHE in privacy-preserving data mining.

\end{document}